# Effects of polariton squeezing on the emission of an atom embedded in a microcavity

Paolo Schwendimann and Antonio Quattropani
Institute of Physics. Ecole Polytechnique Fédérale de Lausanne.
CH 1015 Lausanne-EPFL, Switzerland

## Abstract

We are interested in the emission characteristics of an atom interacting with the polaritonic excitations of a semiconductor or insulator embedded in a one-mode cavity. Since polaritons inside of the cavity exhibit intrinsic squeezing, we discuss how the squeezed polariton ground state modifies the atomic emission and how an indirect evidence of polariton squeezing outside of he cavity may be obtained.

Polaritons are mixed photon-excitations (exciton or phonon) states, which exist in dielectrics and semiconductors [1]. As it is well known, the ground state of the polaritons is a squeezed state and the magnitude of its squeezing depends on the characteristics of the material [2-5]. There is recent interest in engineering the ground state of composite systems consisting of impurities embedded in semiconductors or dielectrics [6]. An atom or an impurity embedded in a polaritonic host medium, offers an interesting example of such an engineering process. The choice of the host material is relevant, because it has been shown that the magnitude of the squeezing of the polariton ground state varies with the choice of the polaritonic material [3]. This characteristic allows optimizing the choice of the host medium in function of the properties of its polaritonic ground state. We expect several characteristics of the radiation field emitted by the composite system to change depending on the magnitude of the ground state squeezing. In particular, the changes in the emission characteristics of such a composite system may allow for indirect evidence of the polariton squeezing outside the medium.

In order to investigate this point one has to choose a configuration of the polaritonic medium, which allows observing the polaritonic characteristics in the emitted radiation field. In a large polaritonic bulk medium, intrinsic squeezing is present [3]. However, a direct observation of polaritons in the emission from such a system is not possible because the translational symmetry of the medium is broken at the boundaries and the different polariton states mix together. The situation changes when considering lower dimensional systems. In a two-dimensional medium like a quantum well, emission from polaritons can be observed directly in the longitudinal direction [7] while the polariton dispersion becomes directly accessible when the quantum well is embedded in a microcavity [8, 9]. However, as it is well known [10], intrinsic polariton squeezing is elusive. It doesn't exist outside the medium as a consequence of energy conservation. At most, one can ask whether an indirect evidence of the intrinsic polariton squeezing outside the polaritonic medium may be found.

In this note, we propose a configuration in which indirect evidence of intrinsic polariton squeezing may be detected. We consider a system consisting of a microcavity containing a polaritonic medium in which an impurity has been embedded. If the impurity is prepared in an excited state, its emission inside the cavity gives rise to polaritons and the quantities characterizing the radiation field in the system contain information both on the impurity and on the polaritons. At this stage, our model consists of a two-level atom embedded in a single mode semiconductor microcavity. The interaction between the two-level system and the cavity mode is described using the Jaynes-Cummings model [11] while the polaritons follow from the well-known Hopfield Hamiltonian [1, 2]. Within this model, we first show by perturbation theory that the spontaneous relaxation rate of the excited atomic state is modified by the presence of the polariton vacuum due to its fraction of photons. In a second step the dynamics of the above system is discussed in the framework of a master equation approach. In the stationary regime and without applying an external field, a small occupation is found in the excited state, which we call level $|2\rangle$. By the term occupation, we express the fact that there is a probability different from zero to find the atom in the excited state. This occupation appears as a consequence of the interaction of the ground state photon field with the embedded atom. Detecting the presence of this small occupation, would allow us to give an indirect evidence of polariton squeezing. In order to probe the occupation of the excited state, we need to introduce a third atomic level called level $|3\rangle$ in the following, and to study the transitions induced by an external pump between the levels $|2\rangle$ and $|3\rangle$ When the level $|2\rangle$ is empty, we don't expect to observe any absorption in the transition $|2\rangle \rightarrow |3\rangle$, while absorption should be present when the level $|2\rangle$ has some degree of occupation. Therefore, in the following we introduce a three-level model, in which the levels $|2\rangle$ and $|3\rangle$ are coupled to a new cavity mode chosen such that its direct interaction with the polaritonic medium is negligible and as well as to an external pump field. We notice, that in this configuration, the emission from the transition $|2\rangle \rightarrow |3\rangle$ inside the microcavity is proportional to the one outside the cavity.

The theoretical model reflecting the considerations outlined above consists of the following elements: We first introduce the Hamiltonian describing the polaritonic system in a one-mode cavity, which is the usual phonon- or exciton-mode coupling Hamiltonian[1], expressed for one mode k

$$\hat{H}_{k,pol} = \hbar\omega_f A_k^{1+} A_k^1 + \hbar\omega_k A_k^{2+} A_k^2 + iC_k \left( A_k^{1+} A_k^2 - A_k^{2+} A_k^1 + A_{-k}^1 A_k^2 - A_k^{2+} A_{-k}^{1+} \right) + $$
$$D_k \left( A_k^{1+} A_k^1 + A_k^1 A_k^{1+} + A_k^{1+} A_{-k}^{1+} + A_{-k}^1 A_k^1 \right), \qquad (1)$$

The operators $A_k^{1+}, A_k^{2+}$ and $A_k^1, A_k^2$ are the photon and excitation creation and annihilation operators respectively. We express the coupling constants appearing in (1) in terms of the material parameters as follows: $C_k = \sqrt{\pi \chi \omega_k / \omega_f \varepsilon_\infty}$, $D_k = C_k^2 / \omega_k$. Here $\varepsilon_\infty$ is the dielectric constant at infinity, $\omega_k = \omega_k(k)$ and $\omega_f = \omega_f(k)$ are the exciton and the photon dispersion respectively and $\chi$ is the linear response constant of the medium. All quantities are normalized with the excitation frequency at $k = 0$ which we denote by $\omega_0$. The Hamiltonian $H_{k,pol}$ is then diagonalized, and in the polariton picture reads

$$H_{k,pol} = \hbar\Omega_k^1 B_k^{1+} B_k^1 + \hbar\Omega_k^2 B_k^{2+} B_k^2 \qquad (2a)$$

$$B_k^1 = W_1 A_k^1 + X_1 A_k^2 + Y_1 A_k^{1+} + Z_1 A_k^{2+} \qquad (2b)$$

$$B_k^2 = X_1 A_k^1 + W_1 A_k^2 + Z_1 A_k^{1+} + Y_1 A_k^{2+} \qquad (2c)$$

.

The Hamiltonian (1) has been simplified by assuming $k = -k$ before diagonalization. The quantities $\Omega_k^1, \Omega_k^2$ appearing in (2) are the eigen-frequencies of the photon and exciton polaritons respectively. The explicit expressions of the coefficients W, X, Y, Z appearing in (2b) and (2c) are found in the papers [2, 3]. These coefficients as well as the polariton frequencies, depend on the characteristics of the polaritonic medium. As discussed in [3], the unitary transformation leading from (1) to the diagonalized form of the polariton Hamiltonian (2) is a product of two phase- and squeezing transformations. Therefore, the ground state of the polaritons is a squeezed state. For later use, we remember that it in term of polaritons operators, the photon operator takes the form [2, 3]

$$A_k^1 = W_1^* B_k^1 + X_1^* B_k^2 - Y_1 B_k^{1+} - Z_1 B_k^{2+}. \tag{3}$$

The characteristics of the polariton ground state have been discussed in detail in the literature[10]. In particular, we remember that the photon content of the polaritonic ground state is different from zero and is expressed as

$$\left\langle A_k^{1+} A_k^1 \right\rangle_{gs} = \left\langle n_k^1 \right\rangle_{gs} = |Y_1|^2 + |Z_1|^2 \tag{4}$$

This quantity plays an important role in the following.
When embedding an atom in a polaritonic system, we expect that the squeezing of the polariton ground state may influence the interaction between the polaritons and the atom. In order to investigate this point, we first consider a two-level atom and in consequence add to (1) the Jaynes-Cummings Hamiltonian

$$H_{J-C} = \hbar \omega_a \sigma^+ \sigma^- + ig \left( A_k^1 \sigma^+ - A_k^{1+} \sigma^- \right), \tag{5}$$

In (5), $\omega_a$ is the frequency difference between the atomic states and the operators $\sigma^+, \sigma^-$ describe the transition between the two atomic levels. The coupling constant $g$ in (5) is the dipole coupling constant. As it is well known, the Hamiltonian (5) is valid when the condition $g < \omega_a$ is satisfied. The photon operators $A_k^1, A_k^{1+}$ are expressed through the polariton operators as shown in (3).

A first indication on the effects of polariton squeezing on the behavior of the two-level atom is obtained within first order perturbation theory. In fact, in this approximation we can show how the presence of a squeezed polariton ground state modifies the transition probability and thus the relaxation rate of the two-level atom. To this end, we consider the resonant case characterized by the choice $\omega_f - \omega_k = 0$ in (1), and calculate the transition probability at resonance between the squeezed ground state of the polaritons with the atom in the excited state, $|i\rangle = |0,0,+\rangle$ and the superposition final state $|fin\rangle = (|1,0,-\rangle + |0,1,-\rangle)/\sqrt{2}$, where the atom is in the ground state and one excitation- or one photon-polariton respectively is present. We obtain

$$\left| \langle fin | H_{J-C} | i \rangle \right|^2 = g^2 \left( |W_1|^2 + |X_1|^2 \right)/2 = g^2 \left( 1 + \left\langle A_k^{1+} A_k^1 \right\rangle_{gs} \right)/2 \tag{6}$$

The effect of the squeezed polariton vacuum is manifest in (6) through the renormalization of the dipole coupling $g$ as a function of the intrinsic photon content of the polariton ground state. Furthermore, since the coefficients W, X, Y, Z, are functions of the material variables $C_k, \omega_f, \omega_k$, the modified transition probability (6) depends on the choice of the polaritonic medium.

Notice that in absence of squeezing i.e. for $\langle A_k^{1\dagger} A_k^1 \rangle_{gs} = 0$ the transition probability becomes equal to $g^2/2$ independently of the choice of the polaritonic material. The value $g^2/2$ of the transition probability in absence of squeezing is a consequence of the final state $|f\rangle$ consisting in the normalized superposition of non-squeezed polariton states and as a matter of fact it expresses the transition probability for an atom coupled to two cavity modes.

In Fig. 1 we present the transition probability (6) calculated for different materials supporting exciton- or phonon-polaritons, which are: CuCl, PbI$_2$, CdSe, GaSe, ZnSe (exciton polaritons) and, CdSe, GaP (phonon polaritons).

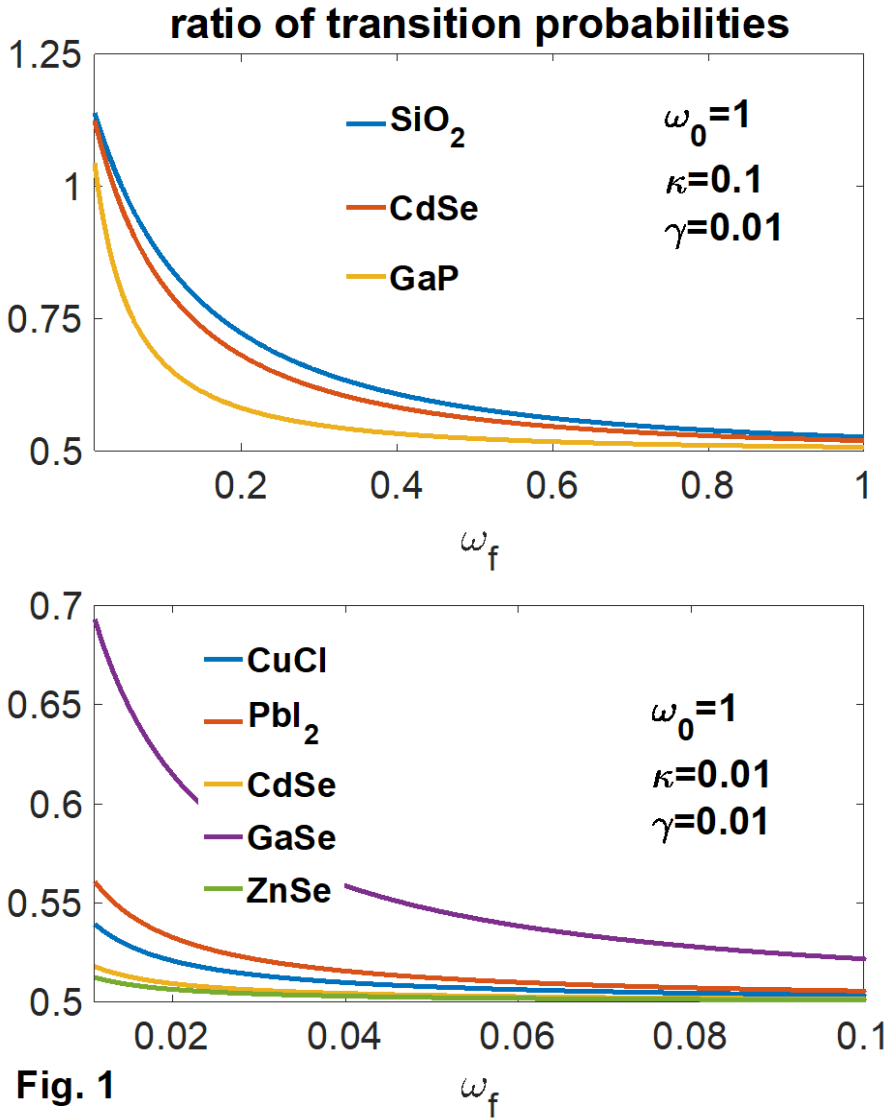

Fig. 1

**Figure 1. Ratio of the transition probabilities at resonance for different materials as a function of the mode frequency $\omega_f$ and the transition probability of an isolated atom:** $\left(1 + \langle A_k^{1\dagger} A_k^1 \rangle_{gs}\right)/2$.

Notice that for an isolated atom, the transition probability between the initial state $|0,+\rangle$ and the final state $|1,-\rangle$ is proportional to $g^2$. Therefore, the quantity $\left(1+\langle A_k^{1\dagger} A_k^1 \rangle_{gs}\right)/2$ in (6), presented in Fig. 1, represents the ratio between the polariton influenced transition probability and the one of an isolated atom. As a consequence, the transition probability for the decay of the excited state of the atom in presence of polariton squeezing is smaller than the one found for the isolated atom, but larger than the one obtained in absence of polaritonic squeezing. This result may be tested experimentally by comparing the relaxation rate of an excited atom embedded in a polaritonic medium with the one obtained for an isolated atom.

The results of Fig. 1 allow us to give a qualitative indication of how engineering of the ground state of the hybrid system consisting of the two-level atom plus polaritonic medium can be realized. It consists in a judicious choice of the material in which the atom or impurity is embedded and of the frequency region in which the system has to be operative. We notice that larger values of polariton squeezing and of the polariton correction to the transition probability may be obtained when choosing a host material exhibiting ultrastrong coupling [12]. We don't enter into the details of the dynamics of a two level atom embedded in a polariton medium. We only notice that, as a consequence of the presence of a squeezed vacuum with a photonic content different from zero, it exists a stationary state of the system also in absence of an external excitation. In this situation, we expect the occupation of the excited atomic level $|2\rangle$, to be different from zero in the stationary state due to the interaction between the atomic levels $|1\rangle$ and $|2\rangle$ and the intrinsic photon field.

From these considerations, it follows, that an indirect evidence of polariton squeezing may be obtained in the following configuration: We consider a three level atom embedded in a cavity. The levels $|1\rangle$ and $|2\rangle$ coincide with the ones of the two-level atom formerly considered and are coupled to the polariton field like in (5), while the transition between the level $|2\rangle$ and the third level is coupled to a different cavity mode, which is strongly out of resonance with respect to the transition $|1\rangle \to |2\rangle$. Assuming that the atom-polariton system is in the stationary regime, we inject into the supplementary mode an external pump field. In presence of a small occupation of level $|2\rangle$, we should observe some absorption in the pump field. The Hamiltonian describing this configuration is

$$H_{total} = H_{k,pol} + \omega_{a2}|2\rangle\langle 2| + (\omega_{a3} - \omega_p)|3\rangle\langle 3| + (\omega_3 - \omega_p)A_3^\dagger A_3 + H_{at-pol}, \qquad (7)$$

with

$$H_{at-pol} = ig_1\left(A_1|2\rangle\langle 1| - A_1^\dagger|1\rangle\langle 2|\right) + ig_2\left(A_3|3\rangle\langle 2| - A_3^\dagger|2\rangle\langle 3|\right) + f\left(A_3 + A_3^\dagger\right) \qquad (8)$$

The interaction-free part (7) consists of the polariton Hamiltonian (2) to which we add the Hamiltonian of a three-level atom. The energies of the levels are expressed through the transition frequencies $\omega_{a1} < \omega_{a2} < \omega_{a3}$ while the non-polaritonic cavity mode is described by the amplitude operators $A_3, A_3^\dagger$ and by the frequency $\omega_3$. The interaction Hamiltonian (8) consists of three terms: the first term describes the coupling of the levels $|1\rangle \to |2\rangle$ the polariton field through (3), while the transition $|2\rangle \to |3\rangle$ is coupled to an external field mode as well as to the non-polaritonic field mode described by the operators $A_3$. Furthermore, we have introduced a coupling of the $|2\rangle \to |3\rangle$ transition to an external classical time-dependent pump field defined as $f(t) = f_0(t)\exp(-i\omega_p t)$ where $f_0(t)$ is a real function. In writing (7) and (8) we have performed the usual unitary transformation into a reference frame rotating with the pump frequency $\omega_p$. Finally, the level $|1\rangle$ has been assumed as reference level.

We discuss the effects of the polariton-atom interaction on the dynamics of the hybrid system working with the Hamiltonians (7) and (8) in a Master Equation approach. In order to derive the master equation including losses, we add to (2) a coupling of the field mode to a bosonic reservoir characterized by the operators $b_q, b_q^\dagger$ and having the form

$$H_{f-res} = \sum_q \omega_q b_q^\dagger b_q +$$
$$\sum_q \left( b_q^\dagger \left( W_1^* B_k^1 + X_1^* B_k^2 - Y_1 B_k^{1+} - Z_1 B_k^{2+} \right) + b_q \left( W_1 B_k^{1+} + X_1 B_k^{2+} - Y_1^* B_k^1 - Z_1^* B_k^2 \right) \right) \qquad (9)$$

The derivation of the Master Equation, describing in the polariton picture the dynamics of the system including the losses of the field mode, requires some care. In fact in the polaritonic strong coupling regime considered in this note, the strong excitation-photon interaction affects the expression of the Lindblad operators describing the losses [13-15]. Therefore, in the Appendix, we present a short derivation of the generalized Lindblad operator, which accounts for the effects of the strong exciton-photon interaction. The Master Equation including the relaxation term derived in the Appendix reads

$$i\hbar \frac{d\rho}{dt} = [H_{total}, \rho] + i\Lambda_f \rho + i\Lambda_{a12}\rho + i\Lambda_{a23}\rho + i\Lambda_{A_3}\rho \qquad (10a)$$

$$\Lambda_f \rho = \left[ F_k(\kappa)\rho, A_k^{1+} \right] + \left[ A_k^1, \rho F_k^+(\kappa) \right] \qquad (10b)$$

$$F_k(\kappa) = \kappa_A W_1^* B_k^1 + \kappa_B X_1^* B_k^2, \qquad (10c)$$

where $A_k^{1+}$ and $A_k$ are given by (3) and the atomic and photon relaxation terms in (10a) have the standart Lindblad form:

$$\Lambda_{a12}\rho = -\gamma_1 \left( |2\rangle\langle 2|\rho + \rho|2\rangle\langle 2| - 2|2\rangle\langle 1|\rho|1\rangle\langle 2| \right) \qquad (10d)$$

$$\Lambda_{a23}\rho = -\gamma_2 \left( |3\rangle\langle 3|\rho + \rho|3\rangle\langle 3| - 2|3\rangle\langle 2|\rho|2\rangle\langle 3| \right). \qquad (10e)$$

$$\Lambda_{A_3}\rho = \kappa_3 \left[ A_3\rho, A_3^+ \right] + \left[ A_3, \rho A_3^+ \right] \qquad (10f)$$

The coefficients $\kappa_A$, $\kappa_B$ and $\kappa_3$ in (10c) and (10f) are the relaxation rates of the polaritons and of the supplementary photon mode respectively and are calculated at the polariton frequencies (see Appendix). Notice that the relaxation rates of phonons or excitons have been assumed to be much shorter than any relaxation rate and have been neglected. When analyzing the Lindblad term in (10b) as expressed in the Appendix, we notice that it contains besides the usual relaxation terms, cross terms between the two polariton operators and their Hermitian conjugates as well as terms containing creation or annihilation operators only. These last terms introduce a phase in the time evolution of the system.

Starting from (10), we can describe the dynamics of the whole system of levels and fields, our aim is to discuss the following situation. Suppose that in a first step the pump is absent. This situation is realized when the initially the system is in the ground state. In this case, inside the cavity the interaction between the polaritons and the transition $|1\rangle \to |2\rangle$ leads to a stationary state characterized by a small fraction of photons and by a small occupation of the level $|2\rangle$ while both the level $|3\rangle$ and the mode $A_3$ are empty. In Fig. 2, we present both the long time evolution of the fraction of photons and the occupation of level $|2\rangle$ showing that indeed a stationary state is obtained for a long enough time.

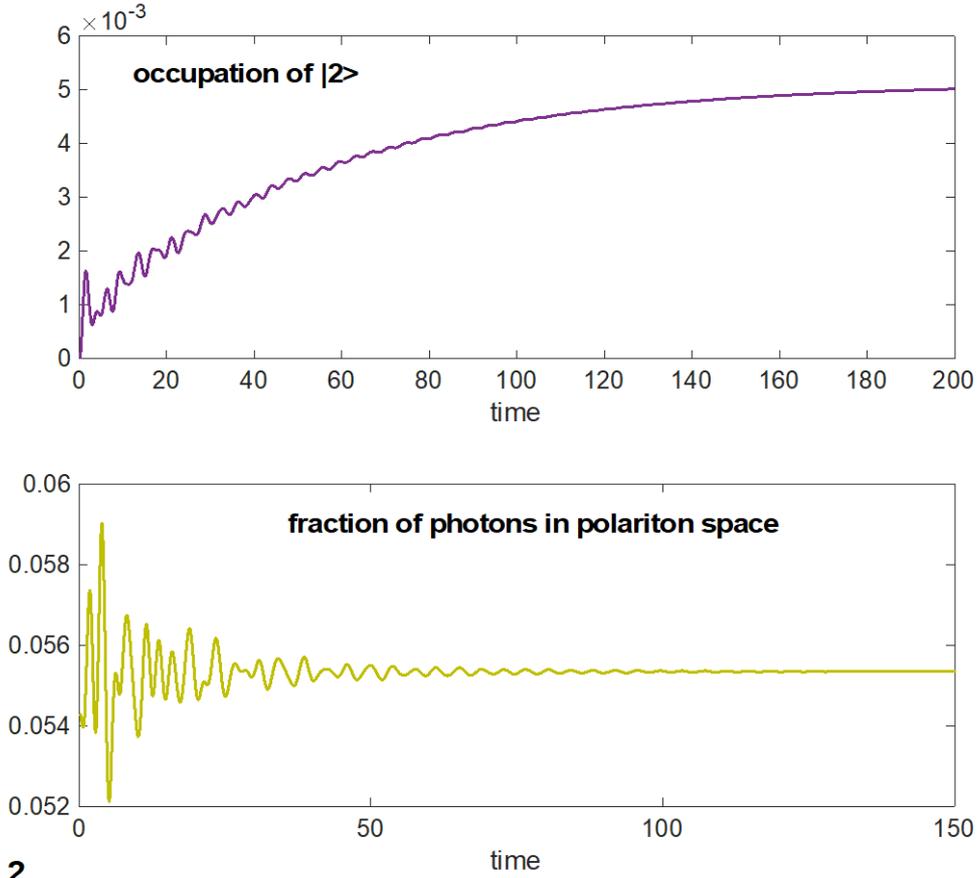

**Fig. 2**

Figure 2. Occupation of the state |2> and fraction of photons in polariton space as a function of time. Material parameters of $SiO_2$ are used and $g_1=0.2$, $g_2=0.3$, $f=1$, $\kappa=0.1$, $\omega_f=1$, $\gamma_1=\gamma_2=0.01$. Initial condition: ground state of the system.

We notice, that in Fig. 2, the stationary photon content is achieved and it is expressed by (4) as expected. We than inject into this stationary system the pump acting on the mode $A_3$ and look for the occupation of the level $|3\rangle$ In order to describe this situation we separate the calculations into two parts: We first solve the master equation (10) for the whole system without pump i.e. $f_0=0$ and we chose the ground state of the system thus obtained as the initial state. In a second step we inject into this stationary system a pump coupled to the mode $A_3$ and discuss the evolution induced by the pump in the stationary system. To this end we need only to consider the evolution described by the master equation

$$\frac{d\rho_1}{dt}=[H_1,\rho_1]+i\Lambda_{A3}\rho_1+i\Lambda_{a23}\rho_1 \qquad (11a)$$

with

$$H_1=\left(\omega_3-\omega_p\right)A_3^\dagger A_3+g_2\left(A_3^\dagger|2\rangle\langle 3|+A_3|3\rangle\langle 2|\right)+\left(\omega_{a3}-\omega_p\right)|3\rangle\langle 3|+f_0(t)\left(A_3+A_3^\dagger\right) \qquad (11b)$$

In the our calculations we have chosen a pump pulse having the form of a Gaussian function

$$f_0(t) = c_0 \exp(-(t-t_0)^2/2\sigma^2)/\sqrt{\pi}\sigma$$

Here $c_0$ is the amplitude of the pump field. We expect that the pump intensity absorbed during excitation shall be re-emitted as a pulse. This behavior is presented in Fig. 3.

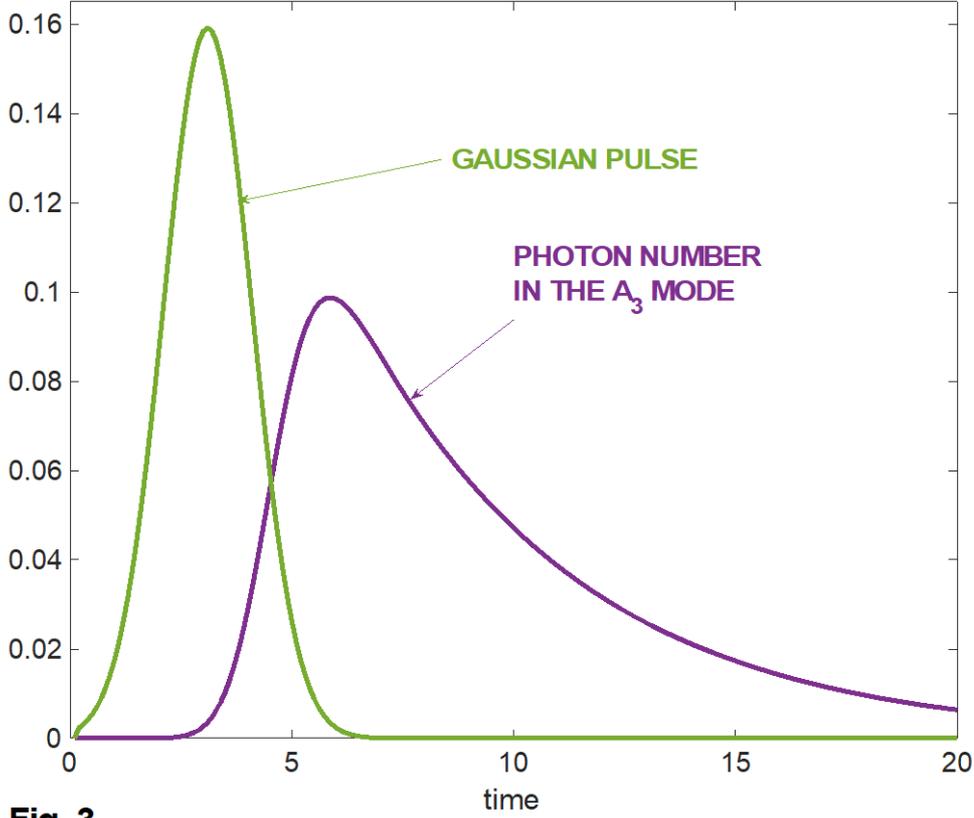

**Fig. 3**

**Figure 3. Photon number in the mode $A_3$ as a function of time under the action of a Gaussian pulse. Material parameters of SiO$_2$ are used and $g_1 = 0.2, g_2 = 0.3, f = 1, \kappa = 0.1, \omega_f = 1, \gamma_1 = \gamma_2 = 0.01$. Initial condition: stationary state**

Schematically, Fig. 3 tells us that during the time evolution, the emission from the transition $|2\rangle \rightarrow |3\rangle$ starts as soon as the maximum of the incoming Gaussian pulse is reached. The delay between the beginning of the action of the pump pulse on the transition and the starting of the emitted pulse is understood in terms of absorption. The incoming pulse needs to be absorbed before the emission can take place. The absorption process takes place because a small occupation is present in level $|2\rangle$ as a consequence of polariton squeezing as shown in Fig.2. This behavior is confirmed by looking at the time evolution of the occupation of levels $|2\rangle$ and $|3\rangle$, which is presented in Fig. 4.

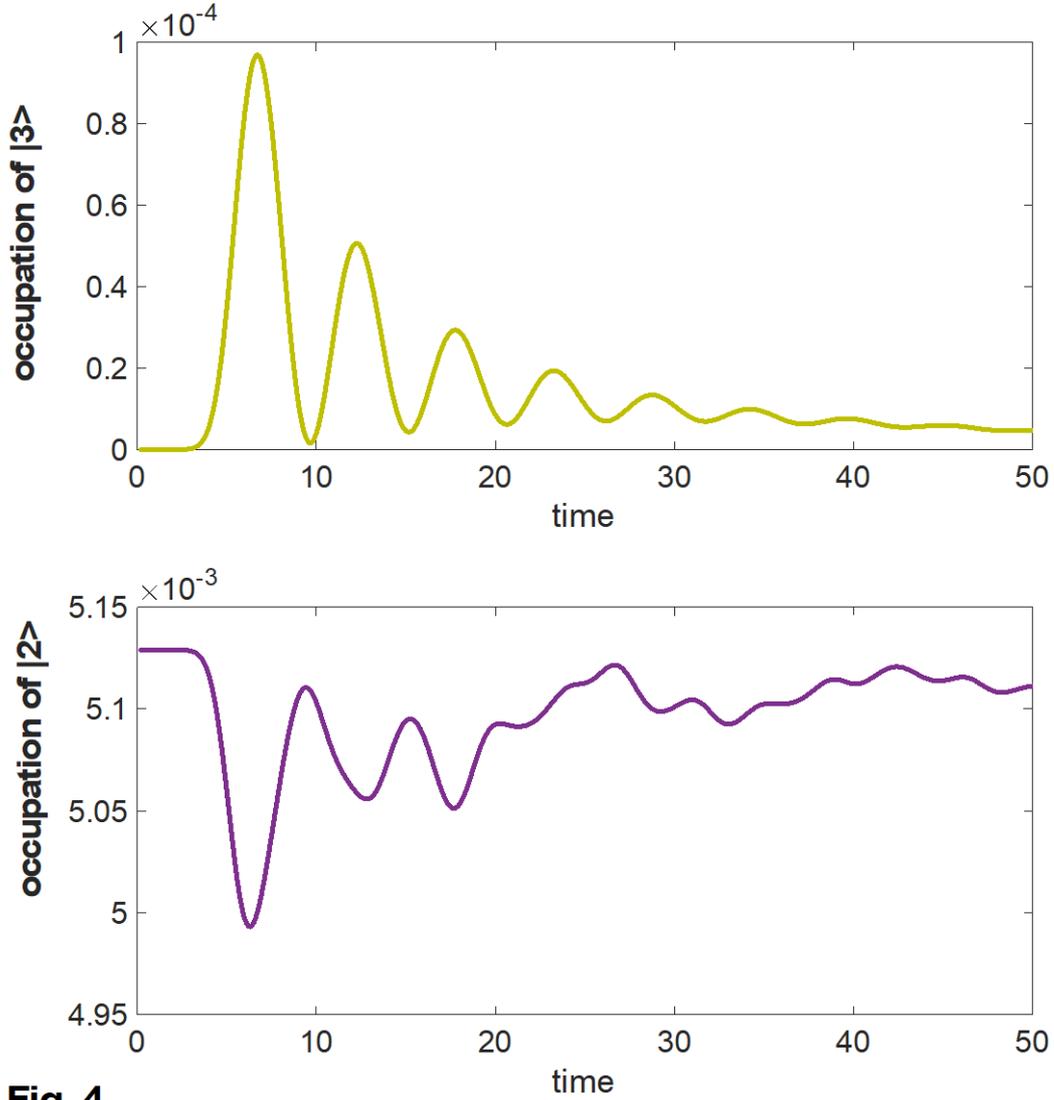

**Fig. 4**

**Figure 4.** Occupation of the states |2> and |3> respectively as a function of time. Material parameters of SiO$_2$ are used and $g_1$ =0.2, $g_2$ =0.3, $f$ =1, $\kappa = 0.1$, $\omega_f = 1$, $\gamma_1 = \gamma_2 = 0.01$. **Initial condition: stationary state**

Starting from their respective initial values, the occupations of levels $|2\rangle$ and $|3\rangle$ have the opposite behavior in time. The occupation of level $|2\rangle$ decreases in time while the one of level $|3\rangle$ grows in the same time interval. This behavior of the occupations indicates that the system is absorbing the pump and generating an occupation in level $|3\rangle$ at the cost of the one of level $|2\rangle$. After the maximum of the absorption is attained, the time dependence of the two levels is interchanged. Comparing the behavior of the occupation to the one of the emission in Fig. 3, we notice that the emission starts becoming relevant before the maximum of the occupation of the level $|3\rangle$ is achieved. Since the occupation of level $|2\rangle$ cannot be measured outside the polaritonic material, we have to extract the indirect information on polariton squeezing from the interaction of the transition $|2\rangle \to |3\rangle$ with the external pump. From the results in Fig.3, we know that an absorption process is taking place between the levels $|2\rangle$ and $|3\rangle$. Therefore, indirect evidence of intrinsic polariton squeezing may be achieved by measuring the absorption in the transition $|2\rangle \to |3\rangle$ in the proposed configuration.

As stated at the beginning of the paper, the goal of our investigation was answering the following question: How does the presence of intrinsic squeezing in the polaritonic medium influence the emission characteristics of the atom plus polaritons system. We have shown that, in presence of intrinsic squeezing the transition probability for the embedded atom becomes smaller than the one of the isolated atom and depends on the magnitude of the intrinsic squeezing and on the material parameters. For the choice of polaritonic materials presented above, the corrections are small but they may become more relevant when choosing materials with larger photon-excitation coupling. A second indirect evidence of polariton squeezing can be found in the following configuration: due to the interaction between the polaritons and the atomic levels, in the stationary state there is a small occupation probability for the excited atomic state (level $|2\rangle$). We add to our model a third level (level $|3\rangle$), which is energetically distant from the polariton state and is coupled it to a corresponding cavity mode. We inject a pump pulse into this new mode and show that absorption between the levels $|2\rangle$ and $|3\rangle$ takes place as a consequence of the fact that level $|2\rangle$ is not empty. Both the modifications in the transition probability and the photon absorption in the transition $|2\rangle \to |3\rangle$ could be tested experimentally.

## Acknowledgements

We would like to thank Vincenzo Savona for illuminating suggestions concerning the three-level model

## Appendix

In this Appendix we derive in the polariton picture the expression for the photon relaxation term in the master equation (10). For simplicity we consider the polaritonic part of (10). The interaction between mode and reservoir is

$$H_{SR} = \left(\Gamma^+ A_k^1 + A_k^{1+}\Gamma\right) + \sum_q \omega_q b_q^+ b_q \qquad (A1)$$

$$\Gamma = \sum_q \lambda_q b_q \qquad (A2)$$

where $\lambda_q$ are the coupling constants characterizing the polariton-reservoir interaction.

The Hamiltonian for describing the dynamics of the polaritons coupled to a photon reservoir is in the polariton basis is

$$H_P = \Omega_A B_k^{1+} B_k^1 + \Omega_B B_k^{2+} B_k^2 + \sum_q \omega_q b_q^+ b_q +$$
$$\Gamma^+ \left(W_1^* B_k^1 + X_1^* B_k^2 - Y_1 B_k^{1+} - Z_1 B_k^2 +\right) + \Gamma\left(W_1 B_k^{1+} + X_1 B_k^{2+} - Y_1^* B_k^1 - Z_1^* B_k^2\right) =$$
$$H_S + H_R + H_{SR} \qquad (A3)$$

In the derivation of the master equation, the contribution of the polariton-reservoir interaction up to second order in the interaction and in the Markov limit is expressed as

$$\Lambda_f \rho_S = Tr_R \int_0^\infty dt' \left[H_{SR}, \left[H_{SR}(t'), \rho_R(0) \otimes \rho_S(t)\right]\right] \qquad (A4a)$$

$$H_{SR}(t') = \exp(i(H_S + H_R)t') H_{SR} \exp(-i(H_S + H_R)t') \qquad (A4b)$$

$$\rho_S = Tr_R \rho \qquad (A4c)$$

We consider the case $\rho_R(0) = |0\rangle_{RR}\langle 0|$. In this case when calculating the trace over the reservoir variables only terms $\langle \Gamma \Gamma^+ \rangle$ are different from zero. For simplicity we consider the following two terms in the development of the double commutator, originating from $H_{SR}^{res}$ and $H_{SR}^{anti-res}$ respectively

$$\int_0^\infty dt' \langle \Gamma\Gamma^+(t')\rangle \left[B_k^{1+} B_k^1(t')\rho_S(t) + B_k^1 B_k^{1+}(t')\rho_S(t)\right] =$$
$$\int_0^\infty dt' \sum_q \lambda_q^2 \exp(i\omega_q t')\left[B_k^{1+} B_k^1 \exp(-i\Omega^1 t')\rho_S(t) + B_k^1 B_k^{1+} \exp(i\Omega^1 t')\rho_S(t)\right] =$$
$$i\pi \sum_q \lambda_q^2 \left[B_k^{1+} B_k^1 \delta(\omega_q - \Omega^1)\rho_S(t) + B_k^1 B_k^{1+} \delta(\omega_q + \Omega^1)\rho_S(t)\right]. \qquad (A5)$$

The term containing the delta function with the plus sign vanishes because both polaritons and reservoir states have positive energies only. Therefore we get in the continuum limits for the reservoir variables

$$i\pi\int_0^\infty d\omega\lambda^2(\omega)D(\omega)\left(B_k^{1+}B_k^1\delta(\omega_k-\Omega^1)\rho_S(t)\right)=$$

$$i\pi\lambda^2\omega_A D(\Omega^1)\left(B_k^{1+}B_k^1\rho_S(t)\right)=i\kappa_A B_k^{1+}B_k^1\rho_S(t) \quad (A6a)$$

The sames holdes for $B_k^{2+}B_k^2$

$$i\pi\int_0^\infty d\omega\lambda^2(\omega)D(\omega)\left(B_k^{2+}B_k^2\delta(\omega_k-\Omega^2)\rho_S(t)\right)=i\kappa_B B_k^{2+}B_k^2\rho_S(t) \quad (A6b)$$

where $D(\omega)$ is the density of states of the reservoir.
By applying this result consequently in the development of the double commutator we obtain

$$\Lambda_f\rho=\left[\left(\kappa_A W_1^* B_k^1+\kappa_B X_1^* B_k^2\right)\rho, A_k^{1+}\right]+\left[A_k^1, \rho\left(\kappa_A W_1 B_k^{1+}+\kappa_B X_1 B_k^{2+}\right)\right], \quad (A7)$$

where $A_k^{1+}$ and $A_k^{1+}$ are given by (3). Finally inserting the definition

$$F_k(\kappa)=\kappa_A W_1^* B_k^1+\kappa_B X_1^* B_k^2,$$

we rewrite (A7) as

$$\Lambda_f\rho=\left[F_k(\kappa)\rho, A_k^{1+}\right]+\left[A_k^1, \rho F_k^+(\kappa)\right]$$

which is Eq. (10b).

# References

1. J. J. Hopfield, Physical Review **112**, 1555 (1958).
2. A. Quattropani, L. C. Andreani, and F. Bassani, Il Nuovo Cimento D **7**, 55 (1986).
3. P. Schwendimann, A. Quattropani, and Z. Hradil, Il Nuovo Cimento D **15**, 1421 (1993).
4. M. Artoni and J. L. Birman, Physical Review B **44**, 3736 (1991).
5. T. Boulier, M. Bamba, A. Amo, C. Adrados, A. Lemaitre, E. Galopin, I. Sagnes, J. Bloch, C. Ciuti, E. Giacobino, and A. Bramati, Nature Communications **5**, 3260 (2014).
6. S. Zeytinoglu, A. Imamoglu, and S. Huber, Physical Review X **7**, 021041 (2017).
7. F. Tassone, F. Bassani, and L. C. Andreani, Il Nuovo Cimento D **12**, 1673 (1990).
8. L. C. Andreani, in *Proceedings of the International School of Physics "E. Fermi". Course CL: Electron and Photon Confinement in Semiconductor Nanostructures.*, edited by B. Deveaud, A. Quattropani and P. Schwendimann (IOS Press, Amsterdam 2003, p. 105.
9. V. Savona, C. Piermarocchi, A. Quattropani, P. Schwendimann, and F. Tassone, Phase Transitions **68**, 169 (1999).
10. C. Ciuti and I. Carusotto, Physical Review A **74**, 033811 (2006).
11. E. T. Jaynes and F. W. Cummings, Proceedings of the Ieee **51**, 89 (1963).
12. C. Ciuti, G. Bastard, and I. Carusotto, Physical Review B **72**, 115303 (2005).
13. H. J. Carmichael and D. F. Walls, Journal of Physics a-Mathematical and General **6**, 1552 (1973).
14. H. P. Breuer and F. Petruccione, *The Theory of Open Quantum Systems* (Oxford University Press, Oxford, 2002).
15. I. Carusotto, S. De Liberato, D. Gerace, and C. Ciuti, Physical Review A **85**, 023805 (2012).